\documentclass[%
 reprint,
 amsmath,amssymb,
aps,
prb,
]{revtex4-1}

\usepackage{graphicx}
\usepackage{dcolumn}
\usepackage{bm}
\usepackage{booktabs}
\usepackage{array}
\usepackage{textcomp}
\usepackage{gensymb}
\usepackage{placeins}

\usepackage[caption=false]{subfig}

\begin{document}


\title{Phonon Dispersion of Mo-stabilized $\gamma$-U measured using Inelastic X-ray Scattering}

\author{Z. E. Brubaker$^{1}$, S. Ran$^{2,3}$, A. H. Said$^{4}$, M. E. Manley$^{1}$,  P. S{\"o}derlind$^{5}$, D. Rosas$^{5}$, Y. Idell$^{5}$, R. J. Zieve$^{6}$, N. P. Butch$^{2,3}$, and J. R. Jeffries$^{5}$\\
$^{1}$ Oak Ridge National Laboratory, Oak Ridge, Tennessee 37831\\
$^{2}$ NIST Center for Neutron Research, National Institute of Standards and Technology, Gaithersburg, Maryland 20899\\
$^{3}$ Center for Nanophysics and Advanced Materials, Department of Physics, University of Maryland, College Park, Maryland 20742\\
$^{4}$ Advanced Photon Source, Argonne National Laboratory, Lemont, Illinois 60439\\
$^{5}$ Lawrence Livermore National Laboratory, Livermore, California 94550\\
$^{6}$ Physics Department, University of California, Davis, California\\
}
\date{\today}

\begin{abstract}


We have measured the room-temperature phonon spectrum of Mo-stabilized $\gamma-$U. The dispersion curves show unusual softening near the H point, q=[1/2,1/2,1/2], which may derive from the metastability of the $\gamma-$U phase or from strong electron-phonon coupling. Near the zone center, the dispersion curves agree well with theory, though significant differences are observed away from the zone center. The experimental phonon density of states is shifted to higher energy compared to theory and high-temperature neutron scattering. The elastic constants of $\gamma$-UMo are similar to those of body-centered cubic elemental metals. 
\end{abstract}
\pacs{Valid PACS appear here}
\maketitle


\section{Introduction}

The electronic, magnetic, and structural properties of actinide materials are generally thought to derive from the unusual nature of the 5f electrons, which hover between localized and itinerant. \cite{Transuranics} The large number of stable structural phases at ambient pressure -- three phases in U and Np and six distinct structural phases in Pu -- are speculated to be stabilized by unconventional interactions of the 5f electrons, though the exact mechanism is the subject of ongoing research. Elemental U undergoes different structural phase transitions as a function of temperature, including the $\alpha$-$\beta$ transition at 668$^{\circ}$C and the $\beta$-$\gamma$ transition at 775$^{\circ}$C . The body-centered cubic (BCC) $\gamma$-U phase can be stabilized at room temperature by alloying with a variety of transition metals, including Mo and Nb. \cite{UMo_handbook} Of the $\gamma$-U alloys, $\gamma$-UMo shows the best combination of high density and irradiation performance, which makes it a promising candidate as a replacement for High-Enriched Uranium fuel in nuclear reactors. \cite{JOM}

The $\alpha$-, $\beta$-, and $\gamma$-U phases have been the focus of several experimental and theoretical efforts focusing on the temperature-dependence of the phonon spectrum and how it is influenced by electronic contributions. \cite{alpha_dispersion_INS,U_DOS_INS, alpha_U_model, alpha_DOS_INS,alpha_IXS_ILM, alpha_U_Bouchet, HTHP_U, Soderlind_gamma_dispersion} The most detailed studies have been restricted to $\alpha$-U, and indicate that electronic contributions may be responsible for (1) the unusually large thermal softening  and (2) a large contribution to entropy that stabilizes the high temperature $\beta$- and $\gamma$-U structures. $\beta$- and $\gamma$-U, however, have only minimally been studied, and experimental phonon measurements have been limited to inelastic neutron scattering (INS) techniques at elevated temperatures. \cite{U_DOS_INS} Phonon dispersion curves have not been determined to date, and would not only provide valuable insight into the lattice dynamics of $\gamma$-U, but also a highly sought-after comparison for first principles calculations. \cite{HTHP_U,Soderlind_gamma_dispersion} To that end, we have performed room-temperature momentum resolved inelastic x-ray scattering (IXS) measurements on $\gamma$-UMo containing 20 atomic \% Mo. 

In general, the dispersion curves agree well with theoretical predictions at small q values, though significant differences are observed away from the zone center. The phonon density of states has been calculated within the framework of the Born-von Karman force constant model and qualitatively matches that predicted by theory.

\begin{figure*}[!htpb]
	\centering
	\includegraphics[width=\linewidth]{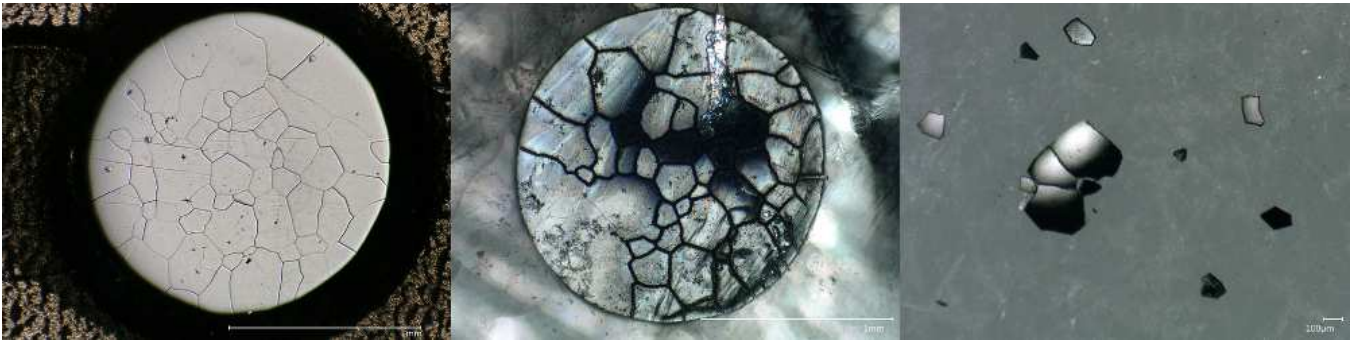}
	\caption{Mo-stabilized $\gamma$-U samples at each stage in the sample preparation. left: After ion beam polishing. center: After laser-cutting. right: After removing single crystals along visible grain boundaries.}
	\label{fig:samples}
\end{figure*}
 
\section{Experimental Methods}

Stoichiometric amounts of depleted uranium and Mo were combined on a water-cooled copper hearth inside of an arc melter to yield a target alloy of U-20 at.\% Mo. The arc melter used four individual arcs with currents up to 30 A to melt the materials, which had a total mass of approximately 13 grams. The combined boule was flipped and re-melted 4 times. After consolidation, a ground, tungsten pull rod was inserted into the molten boule. The pull rod was extracted quickly from the molten boule as in a Czochralski technique in an attempt to quench the high-temperature, body-centered-cubic phase of the U-Mo alloy. The arc melter chamber was evacuated to a pressure of  $<$2e-6 Torr and back-filled with ultra-high purity argon; and, the chamber was continuously gettered during the synthesis and the Czochralski pulling by arc-heating a Zr lump. The as-pulled sample was not a single crystal, but a large-grain polycrystal.

The samples were confirmed to be in the BCC phase via x-ray diffraction (XRD) using the Bruker D8 Discover x-ray diffractometer at 40 kV and 40 mA with copper K-$\alpha$ radiation ($\lambda$=0.15406 nm) that was fitted with the LYNXEYE detector in a symmetric Bragg-Brentano setup over a 2$\theta$ scan range between 20$^{\circ}$ and 90$^{\circ}$. The XRD pattern was analyzed by means of a full Rietveld analysis using GSAS-II and yielded a lattice parameter of a=3.417 \AA\ with a calculated density of $\rho$=17.4 g/cm$^3$. \cite{GSAS} 

Single crystals for IXS were extracted from the polycrystal by polishing and laser-cutting. The polycrystalline samples were mechanically lapped to a thickness of 10 $\mu$m using a precision hand lapping device. Thickness measurements were confirmed using the Zygo NewView 7300 white light interferometer. Once the target thickness was achieved, final polishing was performed using the JEOL ion beam cross section polisher at 6 kV and beam current of 225 $\mu$A at 45 degree tilt for a duration of 30 minutes. The Keyence VHX-6000 was used for optical microscopy imaging, which revealed a mean grain size of 182.72 $\pm$ 29.03 $\mu$m. The samples were then laser-cut along the visible grain boundaries to excise individual grains of the material for use as single-crystals in the inelastic x-ray scattering measurements. The samples at each stage of this process are shown in fig. \ref{fig:samples}.

High energy resolution inelastic x-ray scattering (HERIX) was performed at sector 30 of the Advanced Photon Source using a 23.7 keV incident beam with a 1.5 meV energy resolution. \cite{HERIX1,HERIX2}  Three samples were sandwiched between two diamond windows and the samples were held by graphite paste. A 20-micron thick gold ring was placed around the samples, to ensure that the samples were hermetically sealed and not strained.  A steel sample-holder body clamped the diamond windows and Au foil together, encapsulating the samples for radiological containment necessary for the experiments. 

Throughout this paper, error bars correspond to one standard deviation unless otherwise noted.

\begin{figure*}[!t]
	\centering
	\includegraphics[width=\linewidth]{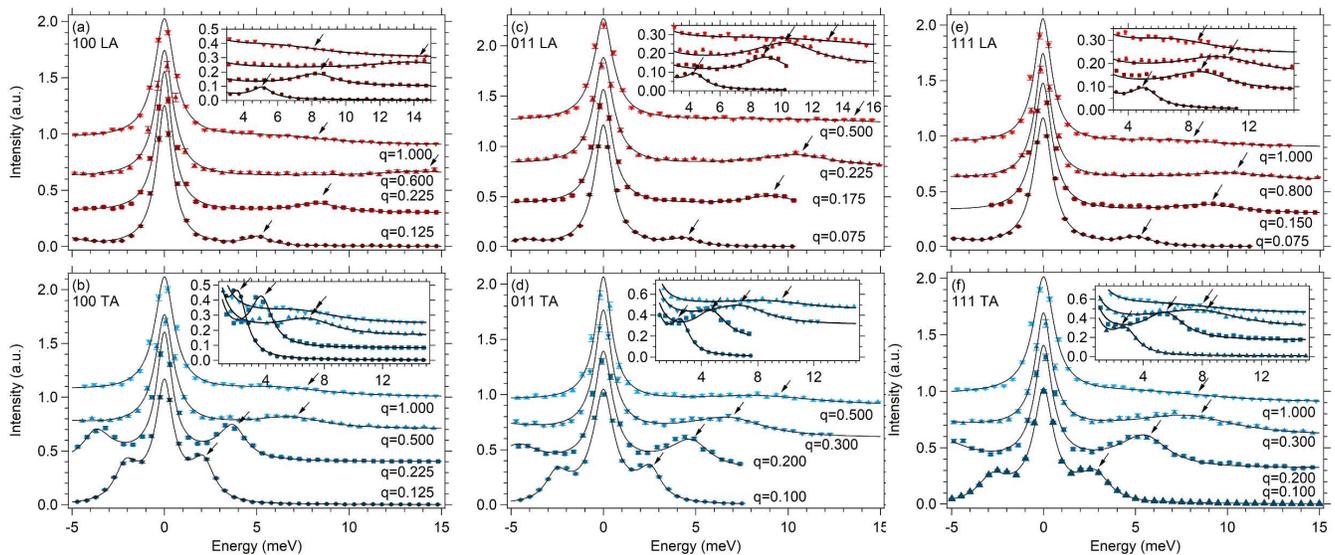}
	\caption{Energy scans at fixed Q along different directions. The symbols represent the raw data and the lines represent the fit along the (a-b) [100] (c-d) [011] and (e-f) [111] directions for different q-values. The top row correspond to the longitudinal acoustic (LA) modes and the bottom row correspond to the transverse acoustic (TA) modes. The peaks are indicated with arrows. Individual curves are offset for clarity. Inset: Energy scan focused on the phonon peaks. Note that the insets use a smaller offset than the main figures.}
	\label{fig:IXS}
\end{figure*}

\section{Results and Discussion}

\subsection{Dispersion Curves}

$\gamma$-UMo forms in the BCC structure with 1 atom per unit cell. As such, 3 acoustic and no optical modes are expected. The transverse acoustic (TA) modes along the high symmetry [100] and [111] directions are degenerate; along the [011] direction, the TA modes polarized along the [01$\overline{1}$] (T$_1$) and [100] (T$_2$) directions are non-degenerate.

Figure \ref{fig:IXS} shows representative inelastic x-ray scattering data obtained for each of the high symmetry directions:  $\overline{\Gamma H}$ ([100]), $\overline{\Gamma N}$ ([011]), and $\overline{\Gamma P}$ ([111]). Only the transverse mode polarized along the [01$\overline{1}$] (T$_1$) direction was measured for the [011] measurements. A typical scan includes a central elastic peak and a Stokes/anti-Stokes pair located on either side of the quasielastic peak. The measured instrumental resolution function was convolved with a Lorentzian to describe the central elastic peak and a damped-harmonic-oscillator (DHO) function to describe the phonon peaks, as described elsewhere and included in the appendix. \cite{DHO, DHO_2017} Close to the zone center, the phonon peaks are readily discernible, but farther from the center, the peaks become challenging to distinguish from the background, likely reflecting some combination of the effects of quenched chemical disorder and the metastability of the cubic phase at room temperature. This is particularly apparent along the [100] and [111] directions, and manifests itself in particularly large linewidths, which are presented in the appendix. 

Figure \ref{fig:dispersion} shows the experimental dispersion curves along the high-symmetry directions. The dispersion curves show a large softening near the H point, which is particularly striking when compared to other BCC elements. Of all the BCC elements, only the group VI metals -- that is Cr, Mo, and W -- and $\beta-$thallium (at 250$\degree$C) show a significant, though smaller, softening near the H point. \cite{Dispersion_Cr,Dispersion_Mo,Dispersion_W,Tl_disp} We speculate that the softening observed in $\gamma-$UMo is related to the metastability of the BCC phase and the strong electron-phonon interactions which have been shown to be unusually strong in uranium.

Figure \ref{fig:dispersion} also shows theoretical predictions for elemental $\gamma$-U using (1) \textit{ab initio} molecular dynamics simulations at 900 K (AIMD; note that we do not compare to the 300 K simulation because it results in divergent diagonal components of the stress tensor,  implying  the phase is  mechanically unstable at 300 K) and (2) self-consistent \textit{ab initio} lattice dynamics (SCAILD) at 1113 K. \cite{HTHP_U, Soderlind_gamma_dispersion} The predicted TA modes for (2) show non-linear curvature near the zone center, which is thought to be due to either (1) the supercell being too small due to computational constraints or (2) the phonon method not fully converging with respect to the anharmonic vibrations. Farther from the $\Gamma$-point, however, the theoretical predictions for the TA modes are expected to hold. 

Because these simulations were performed for elemental $\gamma$-U at elevated temperatures, some differences should be expected when compared to the experimental data at T=300 K for $\gamma$-UMo. By alloying with Mo, the phonon frequency will increase both because the frequency is proportional to M$^{-1/2}$ and because the lattice parameter is reduced by alloying with Mo. The effect of reducing the lattice parameter is similar to that of pressure, and as shown in recent simulations, this will increase the phonon energies, with the most significant deviations occurring at larger q values. \cite{HTHP_U} The effect of temperature is less clear. While the phonon modes in the $\alpha$-U phase experience considerable thermal softening, the $\gamma$-U phase appears rather insensitive to temperature once it becomes mechanically stable, as shown in INS measurements at 1113 K and 1213 K and SCAILD simulations in the 1100 K-1300 K range. \cite{U_DOS_INS, Soderlind_gamma_dispersion} AIMD simulations, on the other hand, show some differences away from the zone center between 900 K and 1300 K. \cite{HTHP_U} Based on these considerations, the experimental and predicted dispersion curves should align well near the zone center, but may differ significantly away from the zone center. Indeed, our results are consistent with this analysis and show good agreement at small q values with both models. Away from the zone center, however, both models predict phonon energies that are significantly lower than measured experimentally. In general, the SCAILD calculations more closely resemble the experimental LA modes, though major differences are observed for the TA modes, partly due to the limitations discussed previously. The AIMD simulations, on the other hand, tend to predict significantly lower phonon energies for both TA and LA modes.

\begin{figure}[!hbtp]
	\centering
	\includegraphics[width=\linewidth]{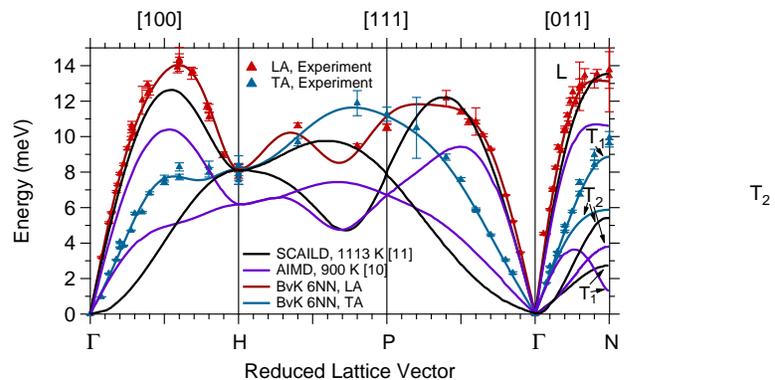}
	\caption{(a) Experimental dispersion curves overlaid with AIMD (purple lines) and SCAILD (black lines) simulations for elemental $\gamma$-U at elevated temperatures and with the results of the BvK model including 6NN (red and blue lines).  \cite{HTHP_U, Soderlind_gamma_dispersion} Near the zone center, the dispersion curves agree well between theory and experiment, though the effects of alloying and temperature likely cause the deviations observed closer to the zone boundaries.}
	\label{fig:dispersion}
\end{figure}

\subsection{Born-von Karman Analysis}

The dispersion curves were analyzed with the Born-von Karman (BvK) force constant model, which is also included in fig. \ref{fig:dispersion}. \cite{BvK} The dispersion curves could be adequately modeled by including 6 nearest neighbors (NN); the results for 4NN and 8NN are included in the appendix. Table \ref{table:AFE_table} lists the atomic force constants calculated from the 6NN fit and also includes the axially symmetric constraints that were used for the 3NN and 4NN atomic force constants. The planar force constants are linear combinations of the atomic force constants and their composition for a BCC crystal can be found in ref \cite{IPC_composition}. The resulting interplanar force constants ($\Phi_n$) are shown in table \ref{table:planar_table}. 

\begin{table}[hbtb]
\centering
\renewcommand\arraystretch{1.2}
\caption{Interatomic force constants (N$\cdot m^{-1}$) calculated from the BvK model for 6NN.}
\label{table:AFE_table}
\begin{ruledtabular}
\begin{tabular}[t]{lcc}
Atomic Position&Force Constant&6NN (N$\cdot m^{-1}$)\\

\hline

$(1,1,1)a/2$&\begin{tabular}{@{}c@{}}1XX \\ 1XY\end{tabular}&\begin{tabular}{@{}c@{}}3.64 \\ 4.76\end{tabular}\\
$(2,0,0)a/2$&\begin{tabular}{@{}c@{}}2XX \\ 2YY\end{tabular}&\begin{tabular}{@{}c@{}}18.88 \\ -3.00\end{tabular}\\
$(2,2,0)a/2$&\begin{tabular}{@{}c@{}}3XX \\ 3ZZ \\ 3XY\end{tabular}&\begin{tabular}{@{}c@{}}1.56 \\ -0.26 \\ 3XX-3ZZ\end{tabular}\\
$(3,1,1)a/2$&\begin{tabular}{@{}c@{}}4XX \\ 4YY \\ 4YZ \\ 4XY\end{tabular}&\begin{tabular}{@{}c@{}}-1.00 \\ 0.36 \\ 1/8*(4XX-4YY) \\ 1/3*(4YZ)\end{tabular}\\
$(2,2,2)a/2$&\begin{tabular}{@{}c@{}}5XX \\ 5XY\end{tabular}&\begin{tabular}{@{}c@{}}1.35 \\ 2.02\end{tabular}\\
$(4,0,0)a/2$&\begin{tabular}{@{}c@{}}6XX \\ 6YY\end{tabular}&\begin{tabular}{@{}c@{}}2.90 \\ -1.48\end{tabular}\\
\end{tabular}
\end{ruledtabular}
\end{table}

\begin{table}[hbtp]
\centering

\renewcommand\arraystretch{1.2}
\caption{Interplanar force constants ($\Phi_n$, N$\cdot m^{-1}$) calculated from the BvK model for 6NN.}
\label{table:planar_table}
\begin{ruledtabular}
\begin{tabular}[t]{lcccccc}
Branch&$\Phi _1$&$\Phi _2$&$\Phi _3$&$\Phi_4$&$\Phi_5$&$\Phi_6$ \\
\hline
$[100]$L&34.80&61.06&$-$7.99&5.81&&\\
$[100]$T&23.95&10.03&2.85&-2.95&&\\ 
$[011]$L&69.20& 20.08 &&&&\\ 
$[011]$T$_1$&31.54 & $-$2.45 &&&&\\
$[011]$T$_2$&13.88& 1.84 & &&& \\
$[111]$L&1.97&25.78&26.54&12.90&-1.70&10.79 \\
$[111]$T&30.89&37.90&-4.08&2.00&-0.01&-1.34\\
\end{tabular}
\end{ruledtabular}
\end{table}

\begin{figure}[hbtp]
	\centering
	\includegraphics[width=\linewidth]{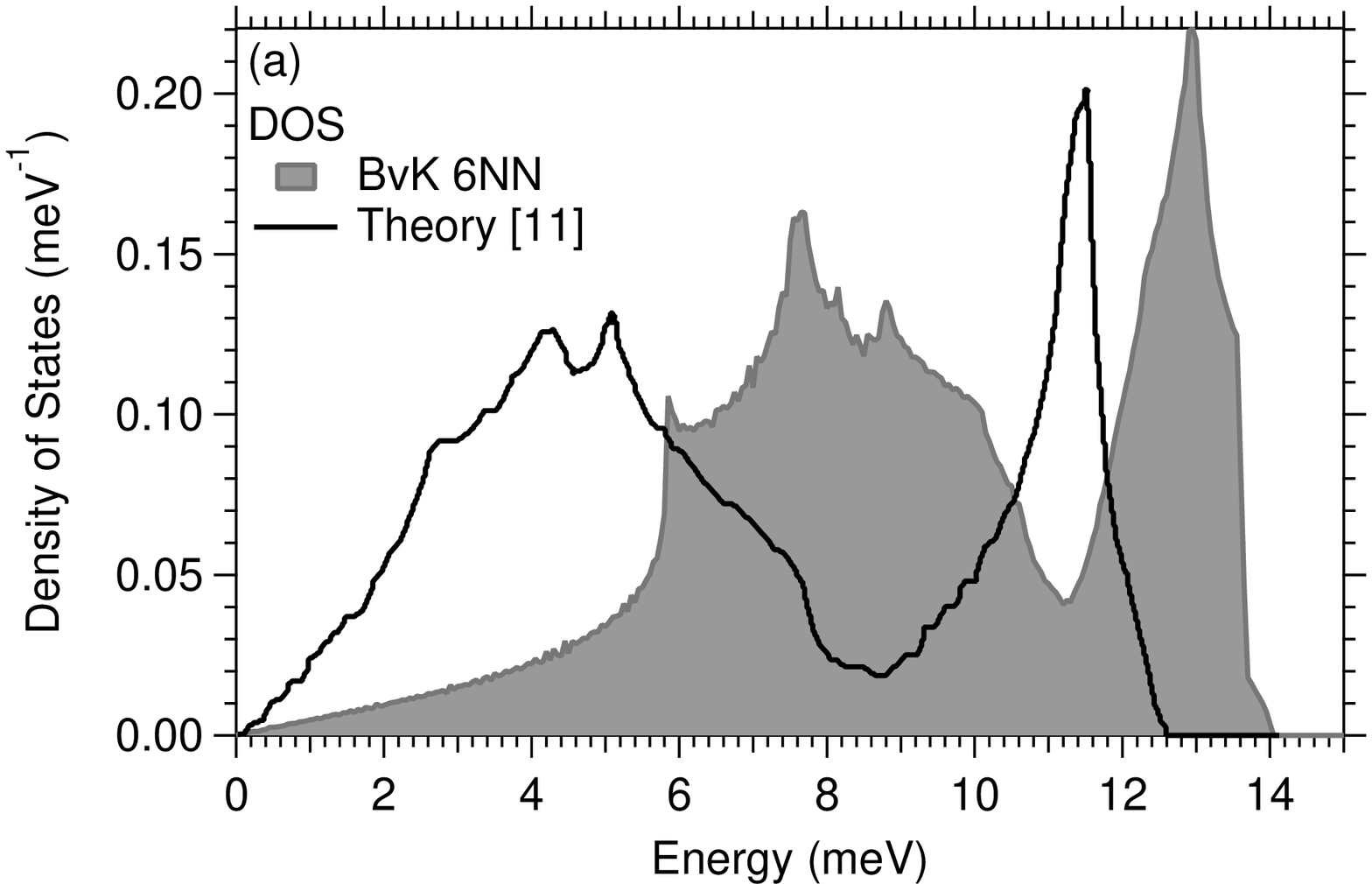}
	\includegraphics[width=\linewidth]{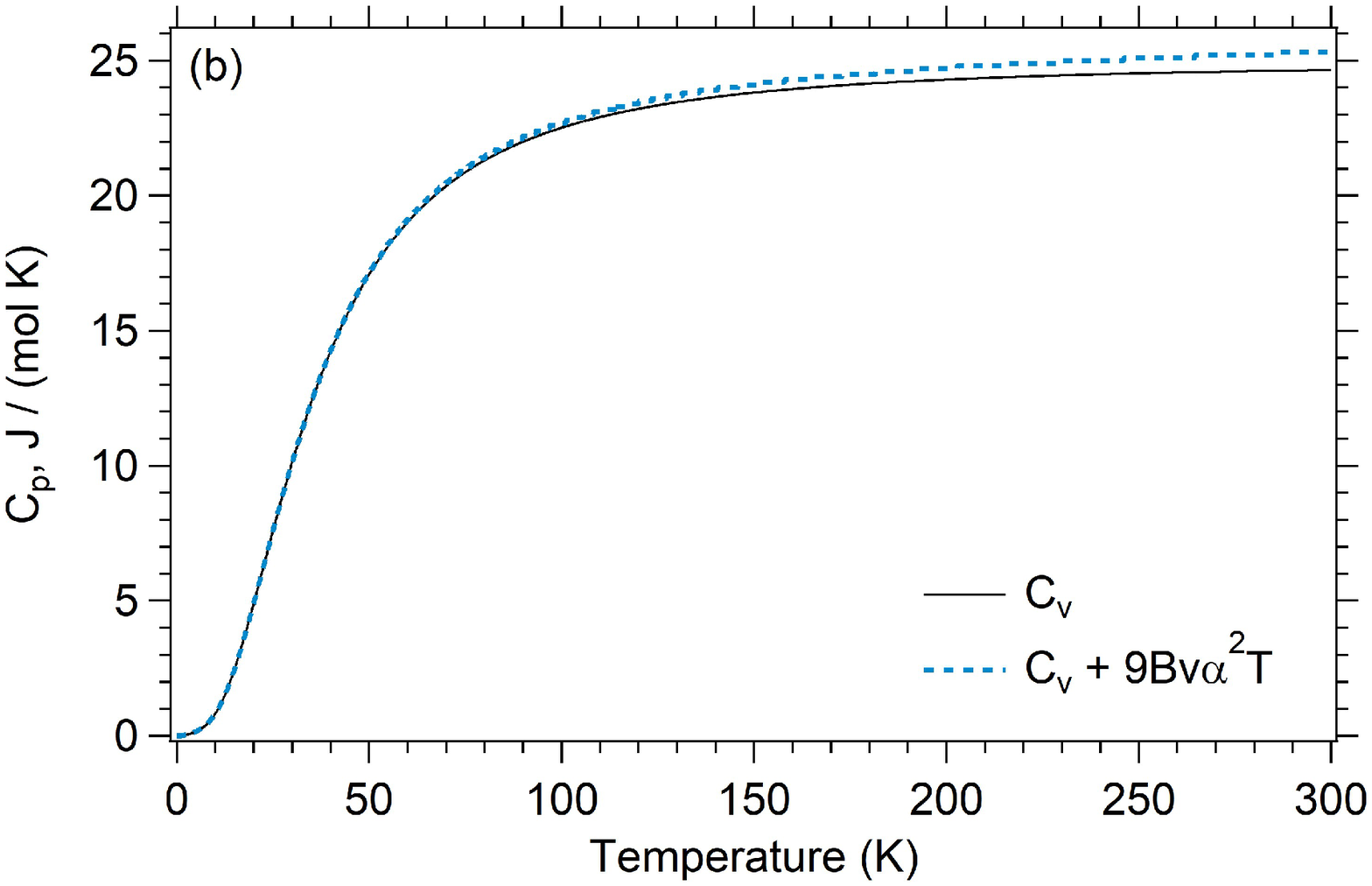}
	\caption{(a) Calculated phonon density of states from BvK analysis (filled gray area) and SCAILD predictions (black line). \cite{Soderlind_gamma_dispersion} Our calculated density of states extends to higher energies than both that determined from SCAILD simulations and that measured via INS at elevated temperatures for $\gamma$-U, likely due to alloying and temperature effects. \cite{Soderlind_gamma_dispersion,U_DOS_INS} The two-peak structure, however, is reproduced in our data. (b) Specific heat calculated from the PDOS.}
	\label{fig:PDOS}
\end{figure}

Using the atomic force constants obtained from the BvK model, the phonon density of states (PDOS) was calculated using a grid size of $b_i=(2 \pi /a)/250$. \cite{dynamic_matrix} Our calculated PDOS (fig. \ref{fig:PDOS}a) shows a similar 2-peak structure as found via INS and theoretical predictions, though the peaks are located at slightly elevated energies. \cite{Soderlind_gamma_dispersion,U_DOS_INS} As discussed previously, these differences are plausibly explained by alloying and temperature effects.

The calculated specific heat is shown in figure \ref{fig:PDOS}b. The lattice specific heat at constant volume can be calculated from the PDOS using 

\begin{equation}
\dfrac{C_v(T)}{3Nk_B} = \int \dfrac{(\beta hv)^2g(v)exp(\beta hv)dv}{(exp(\beta hv)-1)^2},
\end{equation}

where $g(v)$ is the PDOS, $v$ is the frequency, and $\beta=1/(k_BT)$. The contribution to the specific heat from the thermal expansion is given by $C_p-C_v=9B\alpha^2T/\rho$, where B=153 GPa is the bulk modulus calculated using B=1/3$(C_{11} + 2C_{12})$, where the average values of $C_{11}$ and $C_{12}$ from table \ref{table:Elastic_constants_table} were used, $\rho$=17.4 g/$cm^3$ is the density, $\alpha=11.54$x$10^{-6}$/K is the thermal expansion coefficient at T=300 K (from reference \cite{UMo_handbook}) and was assumed to scale as $C_v(T)$ with temperature, T.

\subsection{Elastic Constants}

The sound velocity along each direction can be found by performing a linear fit to the low$-$q region. These velocities are related to the elastic constants via the Green$-$Christoffel equation, and for a cubic system result in only 3 unique elastic constants$:$ C$_{11}$, C$_{12}$, and C$_{44}$. \cite{Elastic_Constants} In table \ref{table:Elastic_constants_table}, we have listed the sound velocities determined for each direction, as well as the elastic constants calculated from (a) only the [100] TA and LA and [011] LA modes, (b) from the [011] TA and [111] LA and TA modes and (c) calculated from \cite{Soderlind_gamma_dispersion}. Curiously, C$_{12}$ differs by about $\sim30$\% depending on which symmetry directions are used for the calculation. 

Table \ref{table:Elastic_constants_table} also includes the elastic constants from most other stable elemental BCC structures (excluding most alkali and alkaline-earth elements). \cite{EC_Li,EC_Ba,EC_V,EC_Cr,EC_Fe,EC_Ti,EC_Zr,EC_Hf,EC_Nb,EC_BCC_Metals,EC_Ce} The $\beta-$phases listed in table \ref{table:Elastic_constants_table} each refer to structures that form the BCC phase at elevated temperatures, as is the case in $\gamma-$U. The elastic constants and anisotropy factor of Mo-stabilized $\gamma-$U show similarities to those determined for the group IV elements -- specifically, $\beta$-Ti, $\beta$-Zr, and $\beta$-Hf. The BCC phases of Ti, Zr, Hf, and U are each thought to arise from phonon-phonon interactions, so it is reasonable to expect similarities in their measured elastic constants. \cite{Soderlind_gamma_dispersion,BCC_group4}

\begin{table*}[hbtp]
\centering
\renewcommand\arraystretch{1.2}
\caption{Sound velocities (m$\cdot$s$^{-1}$) along symmetry directions and elastic constants (GPa) for $\gamma$U-Mo along with most other elemental BCC structures at 300 K. The elastic constants for $\gamma-$UMo were calculated from (a) the [100] TA/LA and [011] LA modes and (b) from the [111] TA/LA and [011] TA modes. The method refers to the measurement technique, which is either inelastic x-ray scattering (IXS), inelastic neutron scattering (INS), ultrasonic, or theoretical calculations.}
\label{table:Elastic_constants_table}
\begin{ruledtabular}
\begin{tabular}[t]{lcccccc}

v$_L$[100]=3295 & \multicolumn{5}{c}{v$_L$[011]=3517}&v$_L$[111]=3260 \\
v$_T$[100]=1507 &\multicolumn{5}{c}{v$_{T1}$[011]=1394}&v$_T$[111]=1425 \\
\hline
System&Method&C$_{11}$&C$_{12}$&C$_{44}$&\begin{tabular}{@{}c@{}}Anisotropy factor \\ 2C$_{44}$/(C$_{11}$-C$_{12}$)\end{tabular}&Reference\\

\hline
$\gamma$-UMo&IXS&189 & 163 & 40 &3.01& This work, (a) \\ 
$\gamma$-UMo&IXS&179 & 111 & 38 &1.14& This work, (b) \\ 
$\gamma$-U (1113 K)&theory&180 & 138 & 21 &1.0& Calculated from \cite{Soderlind_gamma_dispersion}\\ 
\hline
Li&ultrasonic&13.42&11.3&8.89&8.39& \cite{EC_Li}\\
Ba&INS&10&5.42&9.5&4.15& \cite{EC_Ba}\\
V&ultrasonic&230 & 120 &43  &0.78&\cite{EC_V} \\ 
Cr&ultrasonic&348.4 &70.2  &100.7 &0.72&\cite{EC_Cr} \\
Fe&ultrasonic&233.1 & 135.44 &117.8 &2.41&\cite{EC_Fe} \\
$\beta-$Ti (747 K)&INS&134 & 110 &36 &3.0&\cite{EC_Ti} \\ 
$\beta-$Zr (642 K)&INS&104 & 93 &38 &6.91&\cite{EC_Zr} \\ 
$\beta-$Hf (1527 K)&INS&131 & 103 &45 &3.21&\cite{EC_Hf} \\ 
$\beta-$Nb&ultrasonic&245 & 139 &29 &0.55&\cite{EC_Nb} \\ 
Mo&ultrasonic&440 &172  &121  &0.90&\cite{EC_BCC_Metals} \\ 
Ta&ultrasonic&261 &157  &81  &1.56&\cite{EC_BCC_Metals} \\ 
W&ultrasonic &523 & 204 &161  &1.01&\cite{EC_BCC_Metals} \\ 
$\beta-$Tl (523 K)&INS&43 & 39 &12  &6.0&\cite{Tl_disp} \\ 
$\delta-$Ce (1036 K)&INS&22.3 & 17.3 &16.2 &6.48&\cite{EC_Ce} \\ 
\end{tabular}
\end{ruledtabular}
\end{table*}

\section{Conclusion}
We have successfully measured the dispersion relation of Mo-stabilized $\gamma-$U, providing a valuable comparison for theoretical models. The dispersion curves show unusual softening near the H point, possibly indicating the metastability of the structure and the strong electron-phonon coupling.  We have analyzed the dispersion relation following the Born-von Karman force constant model and obtained the interatomic and interplanar force constants, as well as the phonon density of states. The obtained elastic constants agree reasonably well with theoretical predictions for $\gamma$-U, and show similarities to group IV elements which also show BCC phases stabilized by phonon-phonon interactions. High temperature measurements to investigate the evolution of the phonon spectrum with temperature and its role in stabilizing the high-temperature structures would be a valuable addition to the present study.

\section{Acknowledgements}
This work was performed under LDRD (Tracking Code 18-SI-001) and under the auspices of the US Department of Energy by Lawrence Livermore National Laboratory (LLNL) under Contract No. DE-AC52- 07NA27344. This material is based upon work supported by the National Science Foundation under Grant No. NSF DMR-1609855. This research used resources of the Advanced Photon Source, a U.S. Department of Energy (DOE) Office of Science User Facility operated for the DOE Office of Science by Argonne National Laboratory under Contract No. DE-AC02-06CH11357. M.E.M was supported by the Center for Thermal Transport Under Irradiation, an Energy Frontier Research Center funded by the U.S. Department of Energy (DOE), Office of Science, United States, Office of Basic Energy Sciences. This manuscript has been authored by UT-Battelle, LLC, under contract DE-AC05-00OR22725 with the US Department of Energy (DOE). The U.S. government retains and the publisher, by accepting the article for publication, acknowledges that the US government retains a nonexclusive, paid-up, irrevocable, worldwide license to publish or reproduce the published form of this manuscript, or allow others to do so, for US government purposes. DOE will provide public access to these results of federally sponsored research in accordance with the DOE Public Access Plan (http://energy.gov/downloads/doe-public-access-plan).

\section{Appendix}

\begin{figure}[hbpt]
	\centering
	\includegraphics[width=\linewidth]{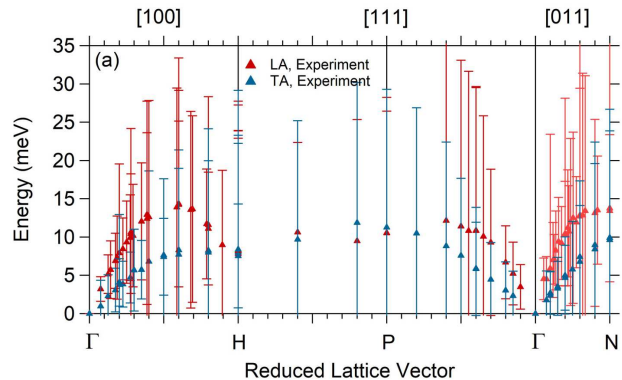}
	\caption{Experimental dispersion curves including the linewidths of phonons.}
	\label{fig:linewidth}
\end{figure}

As mentioned in the main text, the phonon peaks were fit with a damped-harmonic-oscillator (DHO) scattering function convoluted with the instrumental resolution function. The DHO function is given by
\begin{equation}
S(E)=A\dfrac{\Big( \frac{1}{2} \pm \frac{1}{2} + n(|E|) \Big) (\frac{1}{2} \Gamma E)}{(E^2-E_0^2)^{2}+(\frac{1}{2} \Gamma E)^{2})}+B,
\end{equation}
where A is the amplitude, n(E) is the Bose-Einstein distribution function at energy transfer E, $\Gamma$ is the full width at half max, E$_0$ is the phonon energy, and the $\pm$ is determined by E$>$0 (+) or E$<$0 (-). \cite{DHO,DHO_2017} The obtained linewidths are shown in figure \ref{fig:linewidth} and all of the calculated phonon energies and linewidths are collected in table \ref{table:phonons}.

\begin{table*}[hbtp]
\centering
\renewcommand\arraystretch{1.1}
\caption{Phonon energies along symmetry directions and their linewidths ($\Gamma$) at room temperature.}
\label{table:phonons}
\begin{ruledtabular}
\begin{tabular}[t]{lccccccccc}


\multicolumn{9}{c}{LA (q,0,0)} \\

q&E&$\Gamma$&q&E&$\Gamma$&q&E&$\Gamma$ \\

\hline

0	&	0		&				&	0.279	&	10.63	$\pm$	0.23	&	13.53	$\pm$	1.51	&	0.600	&	14.35	$\pm$	0.67	&	19.05	$\pm$	4.23	\\
0.075	&	3.21	$\pm$	0.04	&	1.61	$\pm$	0.22	&	0.279	&	10.45	$\pm$	0.23	&	7.80	$\pm$	1.30	&	0.681	&	13.57	$\pm$	0.34	&	12.84	$\pm$	2.06	\\
0.125	&	5.15	$\pm$	0.05	&	2.53	$\pm$	0.25	&	0.293	&	10.18	$\pm$	0.10	&	6.71	$\pm$	0.60	&	0.694	&	13.65	$\pm$	0.19	&	12.19	$\pm$	1.20	\\
0.143	&	5.73	$\pm$	0.03	&	3.78	$\pm$	0.20	&	0.350	&	12.05	$\pm$	0.79	&	16.14	$\pm$	5.16	&	0.788	&	11.72	$\pm$	0.17	&	7.16	$\pm$	0.99	\\
0.175	&	6.90	$\pm$	0.06	&	3.59	$\pm$	0.31	&	0.386	&	12.94	$\pm$	0.20	&	14.81	$\pm$	1.20	&	0.800	&	11.61	$\pm$	0.22	&	10.83	$\pm$	1.34	\\
0.186	&	7.31	$\pm$	0.06	&	5.44	$\pm$	0.35	&	0.386	&	12.41	$\pm$	0.21	&	11.22	$\pm$	1.28	&	0.800	&	11.11	$\pm$	0.21	&	7.37	$\pm$	1.32	\\
0.200	&	7.92	$\pm$	0.06	&	4.05	$\pm$	0.35	&	0.400	&	12.63	$\pm$	0.28	&	15.21	$\pm$	1.64	&	0.894	&	8.98	$\pm$	0.16	&	9.74	$\pm$	0.99	\\
0.225	&	8.44	$\pm$	0.07	&	5.58	$\pm$	0.46	&	0.587	&	13.89	$\pm$	0.30	&	15.55	$\pm$	1.73	&	1.000	&	7.78	$\pm$	0.24	&	16.72	$\pm$	1.47	\\
0.250	&	9.34	$\pm$	0.10	&	5.38	$\pm$	0.57	&	0.600	&	14.31	$\pm$	0.36	&	11.63	$\pm$	2.99	&	1.000	&	8.22	$\pm$	0.21	&	19.53	$\pm$	1.48	\\
0.275	&	9.91	$\pm$	0.10	&	7.65	$\pm$	0.72	&	0.600	&	14.24	$\pm$	0.24	&	14.92	$\pm$	1.43	&	1.000	&	8.26	$\pm$	0.19	&	18.99	$\pm$	1.50	\\

\hline

\multicolumn{9}{c}{TA (q,0,0)} \\

q&E&$\Gamma$&q&E&$\Gamma$&q&E&$\Gamma$ \\

\hline

0	&	0			&				&	0.275	&	4.71	$\pm$	0.03	&	3.82	$\pm$	0.14	&	0.600	&	8.29	$\pm$	0.23	&	11.82	$\pm$	1.39	\\
0.075	&	0.97	$\pm$	0.01	&	3.39	$\pm$	0.07	&	0.300	&	5.67	$\pm$	0.06	&	5.33	$\pm$	0.34	&	0.800	&	8.25	$\pm$	0.36	&	13.06	$\pm$	2.17	\\
0.125	&	2.29	$\pm$	0.01	&	2.78	$\pm$	0.11	&	0.350	&	5.74	$\pm$	0.05	&	5.01	$\pm$	0.26	&	0.800	&	7.99	$\pm$	0.27	&	11.97	$\pm$	1.70	\\
0.175	&	3.07	$\pm$	0.02	&	2.86	$\pm$	0.11	&	0.400	&	6.81	$\pm$	0.08	&	6.79	$\pm$	0.45	&	1.000	&	7.96	$\pm$	0.31	&	15.88	$\pm$	1.68	\\
0.200	&	3.76	$\pm$	0.02	&	3.00	$\pm$	0.13	&	0.500	&	7.42	$\pm$	0.13	&	9.18	$\pm$	0.75	&	1.000	&	8.12	$\pm$	0.80	&	15.17	$\pm$	4.34	\\
0.200	&	4.06	$\pm$	0.03	&	3.09	$\pm$	0.18	&	0.500	&	7.64	$\pm$	0.19	&	9.96	$\pm$	1.14	&	1.000	&	7.55	$\pm$	0.25	&	14.25	$\pm$	1.52	\\
0.225	&	3.86	$\pm$	0.02	&	3.33	$\pm$	0.12	&	0.600	&	7.73	$\pm$	0.16	&	11.25	$\pm$	0.95	&										\\

\hline

\multicolumn{9}{c}{LA (0,q,q)} \\

q&E&$\Gamma$&q&E&$\Gamma$&q&E&$\Gamma$ \\

\hline

0	&	0			&				&	0.175	&	9.22	$\pm$	0.11	&	5.96	$\pm$	0.74	&	0.300	&	12.84	$\pm$	1.36	&	17.77	$\pm$	6.24	\\
0.056	&	4.54	$\pm$	0.07	&	2.71	$\pm$	0.37	&	0.200	&	10.35	$\pm$	0.12	&	5.95	$\pm$	1.00	&	0.300	&	12.71	$\pm$	0.54	&	16.73	$\pm$	3.27	\\
0.075	&	4.51	$\pm$	0.03	&	2.92	$\pm$	0.17	&	0.200	&	10.44	$\pm$	0.16	&	6.83	$\pm$	1.14	&	0.318	&	12.86	$\pm$	0.40	&	18.54	$\pm$	2.25	\\
0.100	&	5.91	$\pm$	0.04	&	3.25	$\pm$	0.22	&	0.218	&	11.22	$\pm$	0.12	&	7.55	$\pm$	0.68	&	0.335	&	13.44	$\pm$	0.39	&	17.65	$\pm$	2.28	\\
0.118	&	7.00	$\pm$	0.08	&	4.90	$\pm$	0.43	&	0.225	&	10.75	$\pm$	0.14	&	9.62	$\pm$	0.99	&	0.400	&	13.15	$\pm$	0.41	&	12.20	$\pm$	2.58	\\
0.125	&	7.10	$\pm$	0.05	&	4.76	$\pm$	0.39	&	0.236	&	11.97	$\pm$	0.20	&	10.92	$\pm$	1.07	&	0.417	&	13.53	$\pm$	0.26	&	7.07	$\pm$	1.82	\\
0.137	&	8.25	$\pm$	0.08	&	5.14	$\pm$	0.39	&	0.254	&	12.51	$\pm$	0.30	&	11.20	$\pm$	1.68	&	0.500	&	13.75	$\pm$	1.03	&	33.01	$\pm$	5.10	\\
0.155	&	9.41	$\pm$	0.13	&	5.75	$\pm$	0.67	&	0.275	&	11.93	$\pm$	0.22	&	17.53	$\pm$	1.25	&	0.500	&	13.43	$\pm$	2.03	&	25.85	$\pm$	10.73	\\

\hline

\multicolumn{9}{c}{TA (0,q,q)} \\

q&E&$\Gamma$&q&E&$\Gamma$&q&E&$\Gamma$ \\

\hline

0	&	0			&				&	0.150	&	3.31	$\pm$	0.01	&	3.51	$\pm$	0.07	&	0.300	&	6.79	$\pm$	0.13	&	7.33	$\pm$	0.75	\\
0.075	&	1.77	$\pm$	0.01	&	3.81	$\pm$	0.15	&	0.200	&	4.99	$\pm$	0.05	&	5.05	$\pm$	0.33	&	0.400	&	8.42	$\pm$	0.25	&	11.18	$\pm$	1.44	\\
0.100	&	2.70	$\pm$	0.01	&	2.78	$\pm$	0.11	&	0.200	&	4.67	$\pm$	0.05	&	4.23	$\pm$	0.35	&	0.400	&	8.98	$\pm$	0.31	&	13.44	$\pm$	1.75	\\
0.100	&	2.37	$\pm$	0.02	&	3.14	$\pm$	0.23	&	0.250	&	5.79	$\pm$	0.06	&	6.26	$\pm$	0.34	&	0.500	&	9.93	$\pm$	0.37	&	16.77	$\pm$	2.05	\\
0.150	&	3.51	$\pm$	0.02	&	3.82	$\pm$	0.14	&	0.300	&	7.43	$\pm$	0.13	&	9.91	$\pm$	0.73	&	0.500	&	9.63	$\pm$	0.21	&	14.22	$\pm$	1.18	\\

\hline

\multicolumn{9}{c}{LA (q,q,q)} \\

q&E&$\Gamma$&q&E&$\Gamma$&q&E&$\Gamma$ \\

\hline

0	&	0			&				&	0.200	&	10.85	$\pm$	0.77	&	18.80	$\pm$	4.22	&	0.500	&	10.55	$\pm$	0.19	&	15.89	$\pm$	0.98	\\
0.050	&	3.49	$\pm$	0.02	&	2.91	$\pm$	0.23	&	0.200	&	10.82	$\pm$	0.10	&	18.70	$\pm$	0.97	&	0.600	&	9.49	$\pm$	0.10	&	15.83	$\pm$	1.37	\\
0.075	&	5.24	$\pm$	0.02	&	4.10	$\pm$	0.16	&	0.225	&	10.83	$\pm$	0.17	&	20.81	$\pm$	1.47	&	0.800	&	10.63	$\pm$	0.14	&	11.72	$\pm$	1.08	\\
0.100	&	6.71	$\pm$	0.02	&	4.76	$\pm$	0.45	&	0.250	&	11.42	$\pm$	0.15	&	21.68	$\pm$	1.38	&	1.000	&	8.27	$\pm$	0.21	&	14.61	$\pm$	2.33	\\
0.150	&	9.29	$\pm$	0.04	&	9.56	$\pm$	0.42	&	0.300	&	12.16	$\pm$	0.44	&	27.03	$\pm$	3.50	&	1.000	&	8.21	$\pm$	0.09	&	15.64	$\pm$	1.15	\\
0.175	&	10.07	$\pm$	0.08	&	15.74	$\pm$	0.93	&	0.500	&	10.52	$\pm$	0.16	&	17.72	$\pm$	1.01	&	\\									

\hline

\multicolumn{9}{c}{TA (q,q,q)} \\

q&E&$\Gamma$&q&E&$\Gamma$&q&E&$\Gamma$ \\

\hline

0	&	0			&				&	0.200	&	5.91	$\pm$	0.07	&	7.97	$\pm$	0.47	&	0.500	&	11.26	$\pm$	0.40	&	18.04	$\pm$	1.54	\\
0.075	&	2.32	$\pm$	0.06	&	3.22	$\pm$	0.16	&	0.250	&	7.59	$\pm$	0.10	&	10.07	$\pm$	0.47	&	0.600	&	11.89	$\pm$	0.71	&	18.31	$\pm$	1.86	\\
0.100	&	3.06	$\pm$	0.04	&	3.70	$\pm$	0.14	&	0.300	&	8.85	$\pm$	0.17	&	13.56	$\pm$	0.79	&	0.800	&	9.69	$\pm$	0.21	&	15.51	$\pm$	1.28	\\
0.150	&	4.47	$\pm$	0.05	&	4.81	$\pm$	0.20	&	0.400	&	10.50	$\pm$	1.72	&	16.41	$\pm$	3.03	&	1.000	&	8.38	$\pm$	0.09	&	20.77	$\pm$	1.14	\\
0.200	&	5.83	$\pm$	0.06	&	6.08	$\pm$	0.28	&		&				&				&		&				&				\\

\end{tabular}
\end{ruledtabular}
\end{table*}

\begin{figure}[hbtp]
	\centering
	\includegraphics[width=\linewidth]{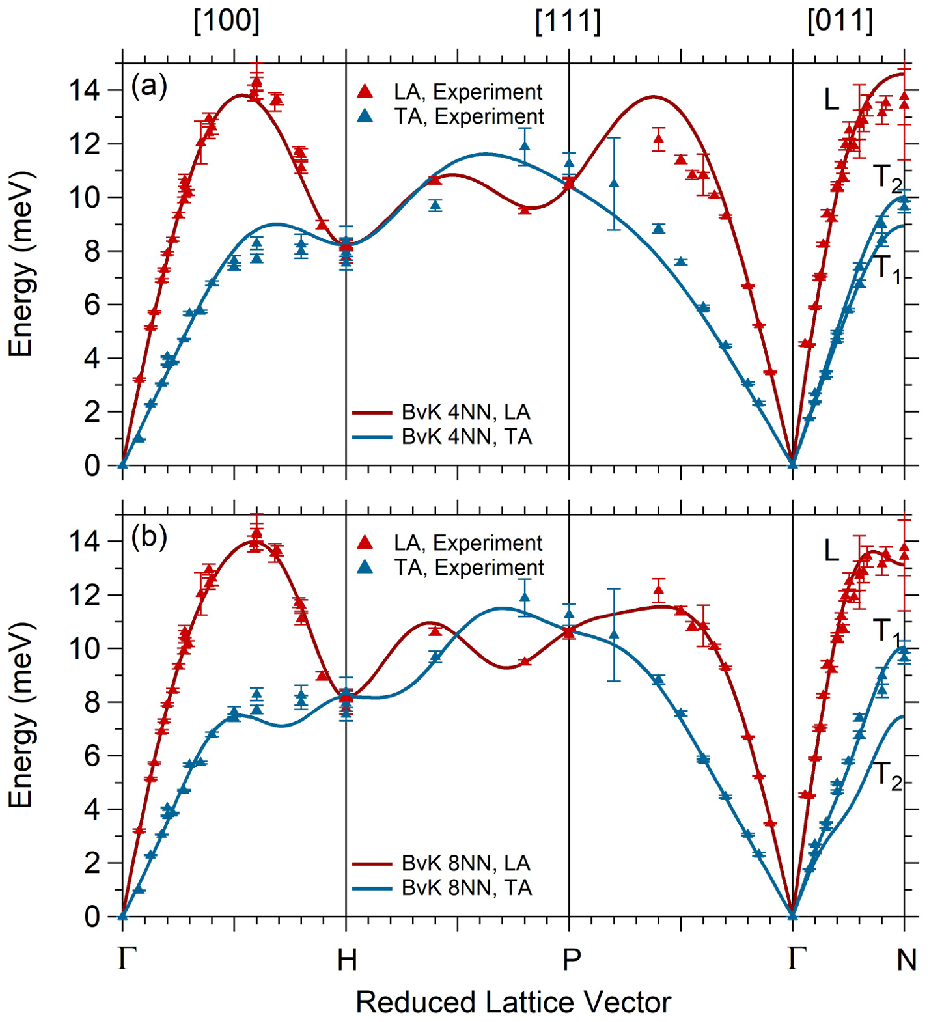}
	\caption{Calculated dispersion curves via BvK analysis using (a) 4NN and (b) 8NN. Major differences are seen in the zone boundary modes between H and P where the data are more sparse and provide fewer constraints.}
	\label{fig:BvK_appendix}
\end{figure}

Figure \ref{fig:BvK_appendix} shows the results of the BvK analysis for 4NN and 8NN. The most notable differences are found in the zone boundary modes between the H and P points, where the data is more sparse and provides fewer constraints. In addition, the 4NN fit also struggles with the [011] LA modes, while the 8NN fit results in additional oscillations.

\FloatBarrier

\end{document}